\documentclass[]{emulateapj}
\usepackage{apjfonts}
\begin{document}

\title{Imaging an Event Horizon: Mitigation of Scattering Toward
  Sagittarius~A*}
\author{Vincent L.\ Fish\altaffilmark{1},
        Michael D.\ Johnson\altaffilmark{2},
        Ru-Sen Lu\altaffilmark{1}, 
        Sheperd S.\ Doeleman\altaffilmark{1,2},
        Katherine L.\ Bouman\altaffilmark{3},
        Daniel Zoran\altaffilmark{3},
        William T.\ Freeman\altaffilmark{3},
        Dimitrios Psaltis\altaffilmark{4},
        Ramesh Narayan\altaffilmark{2},
        Victor Pankratius\altaffilmark{1},
        Avery E.\ Broderick\altaffilmark{5},
        Carl R.\ Gwinn\altaffilmark{6},
        \& Laura E.\ Vertatschitsch\altaffilmark{2}}
\altaffiltext{1}{Massachusetts Institute of Technology, Haystack
  Observatory, Route 40, Westford, MA 01886, USA}
\altaffiltext{2}{Harvard-Smithsonian Center for Astrophysics, 60
  Garden Street, Cambridge, MA 02138, USA}
\altaffiltext{3}{Massachusetts Institute of Technology, Computer
  Science and Artificial Intelligence Laboratory, 32 Vassar Street,
  Cambridge, MA 02139, USA}
\altaffiltext{4}{Astronomy and Physics Departments, University of
  Arizona, 933 North Cherry Street, Tucson, AZ 85721, USA}
\altaffiltext{5}{Perimeter Institute for Theoretical Physics, 31
  Caroline Street North, Waterloo, ON N2L 2Y5, Canada}
\altaffiltext{6}{Department of Physics, University of California,
  Santa Barbara, CA 93106, USA}
\shortauthors{Fish et al}
\shorttitle{Mitigation of Scattering Toward Sgr~A*}
\journalinfo{Accepted to Astrophysical Journal}
\slugcomment{Accepted date: 15 September 2014}
\begin{abstract}
The image of the emission surrounding the black hole in the center of
the Milky Way is predicted to exhibit the imprint of general
relativistic (GR) effects, including the existence of a shadow feature
and a photon ring of diameter ${\sim}50~\mu$as.  Structure on these
scales can be resolved by millimeter-wavelength very long baseline
interferometry (VLBI).  However, strong-field GR features of interest
will be blurred at $\lambda \ge 1.3$~mm due to scattering by
interstellar electrons.  The scattering properties are well understood
over most of the relevant range of baseline lengths, suggesting that
the scattering may be (mostly) invertible.  We simulate observations
of a model image of Sgr~A* and demonstrate that the effects of
scattering can indeed be mitigated by correcting the visibilities
before reconstructing the image.  This technique is also applicable to
Sgr~A* at longer wavelengths.
\end{abstract}
\keywords{galaxies: individual (Sgr~A*) --- Galaxy: center ---
  scattering --- techniques: image processing --- techniques:
  interferometric}

\section{Introduction}

The black hole in the center of the Milky Way, Sagittarius~A*
(Sgr~A*), is the best candidate for spatially resolving a black hole
image at horizon scales.  With a mass of $\sim 4.3 \times 10^6~M_\sun$
at a distance of $\sim 8.3$~kpc
\citep{ghez2008,gillessen2009a,gillessen2009b}, one Schwarzschild
radius subtends ${\sim}10~\mu$as as viewed from the Earth.  When
illuminated by the hot material that surrounds it, general relativity
(GR) predicts that a distant observer would see a bright photon ring
enclosing a darker shadow region whose diameter is ${\sim}50~\mu$as
\citep{bardeen1973,falcke2000}.  Though very small, this size scale is
accessible to ground-based very long baseline interferometry (VLBI) at
$\lambda \le 1.3$~mm.  Sgr~A* is one of the main targets of the Event
Horizon Telescope (EHT), an international collaboration to spatially
resolve and image the GR-dominated innermost accretion and outflow
region around nearby supermassive black holes \citep{whitepaper}.

There have been three motivations for the push toward short-wavelength
VLBI of Sgr~A*.  First, the angular resolution of an interferometric
baseline is $\lambda/|\mathbf{b}|$, where $|\mathbf{b}|$ is the
projected baseline length perpendicular to the line of sight.  The
baseline length is limited by the size of the Earth for a ground-based
array.  At 1.3~mm, the longest EHT baselines will provide an angular
resolution of $\lesssim 25~\mu$as.  Increased angular resolution can
be obtained by observing at shorter wavelengths.  Second, the inferred
intrinsic size of the emission from Sgr~A* \citep[37~$\mu$as at
  1.3~mm;][]{doeleman2008} is larger at longer wavelengths
\citep{doeleman2001,bower2004,shen2005}, indicating that the observed
emission is optically thick at longer wavelengths, obscuring the
shadow near the black hole.  Third and most problematically,
interstellar scattering by free electrons blurs the image of Sgr~A*,
causing its \emph{apparent} size to be proportional to $\lambda^2$,
with the inferred diffractive scale of the scattering corresponding to
baseline lengths of approximately $4500 \times 9300$~km along the
major and minor axes at $\lambda = 1.3$~mm and $1700 \times 3500$~km
at $\lambda = 3.5$~mm
\citep{lo1998,bower2004,bower2006,shen2005,falcke2009,lu2011,akiyama2013}.
This effect dominates the size measurement at $\lambda \gtrsim 3.5$~mm
and is large enough to produce significant blurring even at 1.3~mm
(where a point source would be scattered to be $\sim 22~\mu$as in the
long direction).

The sensitivity and baseline coverage of the EHT will increase
dramatically over the next few years, especially with the inclusion of
phased ALMA (Atacama Large Millimeter/submillimeter Array) as a VLBI
station \citep{fish2013}, allowing EHT targets to be imaged.
Simulated EHT data of the black hole in the nearby giant elliptical
galaxy M87 have demonstrated that the array will be capable of imaging
nearby black holes with sufficient fidelity to resolve the black hole
shadows \citep{lu2014}.  The size of the shadow in M87 is slightly
smaller than in Sgr~A*, but M87 is not significantly
scatter-broadened.  An important question that remains for Sgr~A* is
whether the fine details of the image will be irreversibly washed out
by interstellar scattering \citep[e.g.,][]{broderick2011,yan2014}.  In
this Letter we demonstrate that the effects of interstellar scattering
will be largely invertible in the case of 1.3~mm VLBI of Sgr~A*,
allowing most features of the intrinsic structure to be recovered.

\section{Scatter Broadening} \label{broadening}

\subsection{Theory}

Variations in the density of the tenuous interstellar plasma scatter
radio waves, resulting in temporal and angular broadening of sources
as well as scintillation in frequency and time \citep[see,
  e.g.,][]{rickett1990}.  These variations impart a stochastic phase,
$\phi(\mathbf{r})$, proportional to frequency and density variations,
that changes across the scattering disk.  Here, $\mathbf{r}$ is a
transverse coordinate at the scattering screen.  The variations are
typically quantified by the phase structure function
\begin{equation}
D_{\phi}(\mathbf{r}) = \left \langle \left[ \phi(\mathbf{r}^\prime +
  \mathbf{r}) - \phi(\mathbf{r}^\prime) \right]^2 \right \rangle .
\end{equation}
This structure function exhibits a power-law $D_\phi(\mathbf{r})
\propto |\mathbf{r}|^\alpha$ over scales from 1000 km to over 1000 AU,
suggesting a turbulent cascade \citep[e.g.,][]{armstrong1995}.
Strongly scattered sources often have a power-law index $\alpha$ close
to 2 (the Kolmogorov index is $5/3$), which could arise from
scattering by a medium consisting of discrete scatterers with abrupt
boundaries or from scattering at wavelengths such that the phase
coherence length $r_0 \propto \lambda^{-2/\alpha}$ on the scattering
screen, defined such that $D_{\phi}(\mathbf{r}) \equiv 1$ for
$|\mathbf{r}| = r_0$, is shorter than $r_\mathrm{in}$, the
dissipation scale of the turbulent cascade
\citep{tatarskii1971,lambert1999}.

The dominant effects of scattering depend on the size of the source as
well as the time and frequency resolution with which the source is
observed \citep[e.g.,][]{narayan1989,goodman1989}.  In the
\emph{snapshot-image} regime, observations of a very compact source
with very high time and frequency resolution will detect stochastic
variations in frequency and time due to diffractive scintillation.  As
the integration time or observed bandwidth is increased, the observing
array effectively averages over multiple snapshot images.  In this
\emph{average-image} regime, fast diffractive scintillation is
suppressed, but visibilities fluctuate on significantly broader scales
in frequency and time due to refractive scintillation.  For still
longer integration times or wider bandwidths, an interferometer
averages over many realizations of the scattering screen.  In this
\emph{ensemble-average} regime, the response of an interferometer to a
point source is
\begin{equation}
\tilde{I}(\mathbf{u}) = \exp\left[-\frac{1}{2} D_{\phi} \left(
\frac{\lambda\mathbf{u}}{1 + M} \right) \right],
\label{ideal-equation}
\end{equation}
where the tilde denotes quantities in the Fourier (visibility) domain,
$\mathbf{u} \equiv (u,v)$ represents the projected baseline coordinates
in units of the observing wavelength $\lambda$, and $M$ is the
magnification factor of the scattering screen (the observer-scatterer
distance divided by the source-scatterer distance).

Interstellar scattering is significant throughout the Galactic Center
region \citep[e.g.,][]{vanlangevelde1992}.  The consistency of angular
broadening measurements with $\alpha \approx 2$ implies that
ensemble-average scatter broadening is well approximated by a
Gaussian.  The scattering disk is anisotropic, as is typical for many
lines of sight, which may indicate elongation of turbulent eddies
along their local magnetic field \citep{goldreich1995}.

The ideal ensemble-average scattering kernel
(eq.~(\ref{ideal-equation})) is deterministic and purely real-valued.
Departures of the scattering response from this ideal case can arise
from diffractive or refractive effects.  An extended source suppresses
these types of scintillation noise \citep{gwinn1991,gwinn1998}.  For
Sgr~A*, where the intrinsic 1.3~mm source size is much larger than the
diffractive scale, the diffractive noise is negligible.  Moreover,
because the size of a scatter-broadened point source is smaller than
that of the intrinsic structure at 1.3~mm, the refractive noise is
also partially quenched.

\subsection{Inversion of the Scattering Kernel}

By the van Cittert--Zernike theorem, the visibilities measured by
interferometry are related to the Fourier transform of the sky image
as
\begin{equation}
\tilde{I}(\mathbf{u}) = \int d^2\mathbf{x} \, I(\mathbf{x}) \, e^{-2\pi i
  \mathbf{u} \cdot \mathbf{x}}.
\end{equation}
Convolution in the image domain is equivalent to
multiplication in the visibility domain by the Fourier conjugate of
the convolution kernel: $I(\mathbf{x}) \ast G(\mathbf{x})
\Leftrightarrow \tilde{I}(\mathbf{u}) \, \tilde{G}(\mathbf{u})$, where
$\ast$ denotes convolution.  The Fourier conjugate of the elliptical
Gaussian scattering kernel $G(\mathbf{x})$ in the image domain is an
elliptical Gaussian $\tilde{G}(\mathbf{u})$ in the visibility domain.
Importantly, the scattering kernel is real ($\tilde{G}(\mathbf{u}) \in
\mathbb{R}^+$ for all $\mathbf{u}$) and decreases monotonically in all
directions.  The net effect is that long-baseline amplitudes of the
scattered image are lower than would be measured for the unscattered
image, but visibility phases are unaffected.

\begin{figure}
\resizebox{\hsize}{!}{\includegraphics{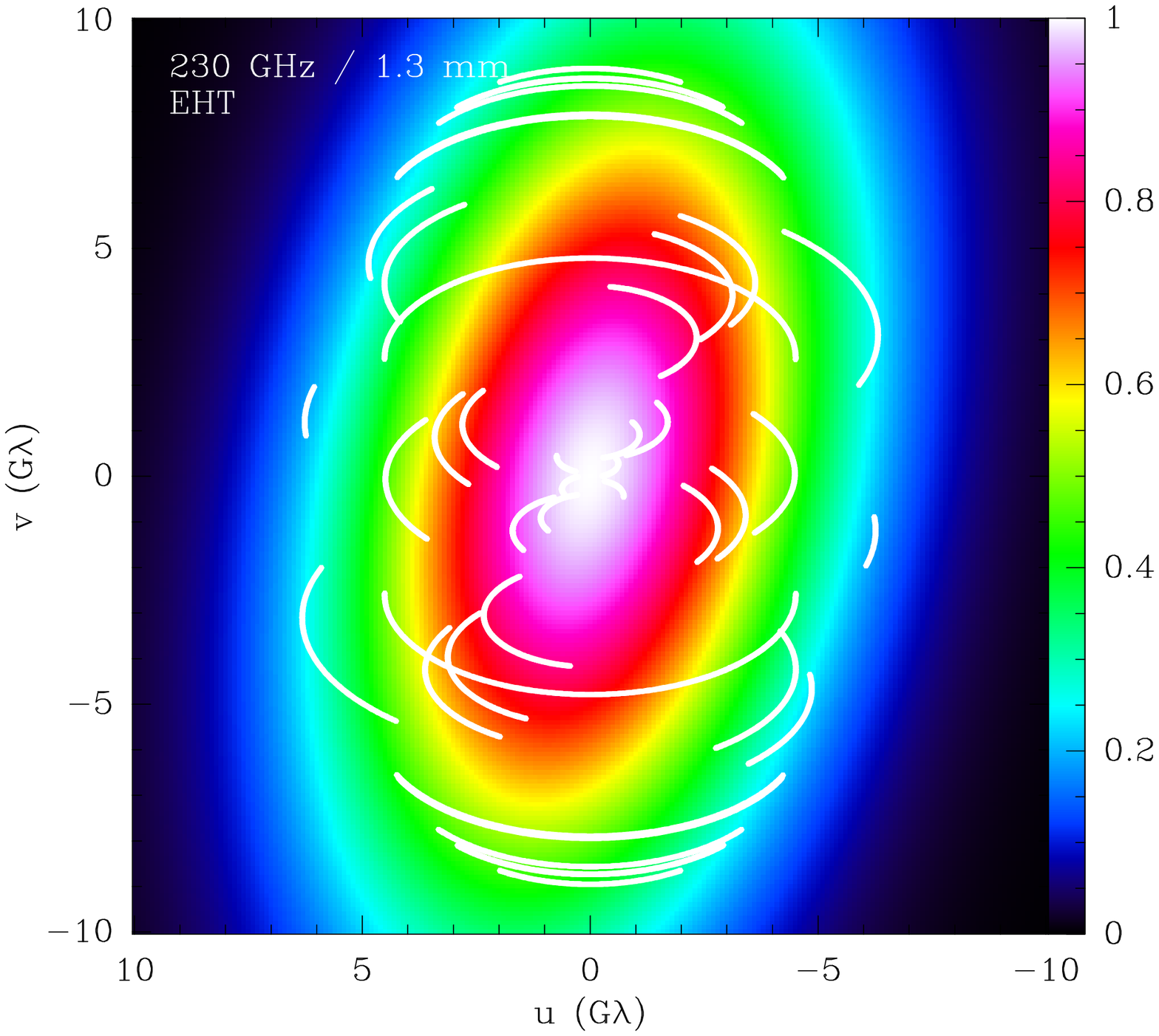}}\\
\resizebox{\hsize}{!}{\includegraphics{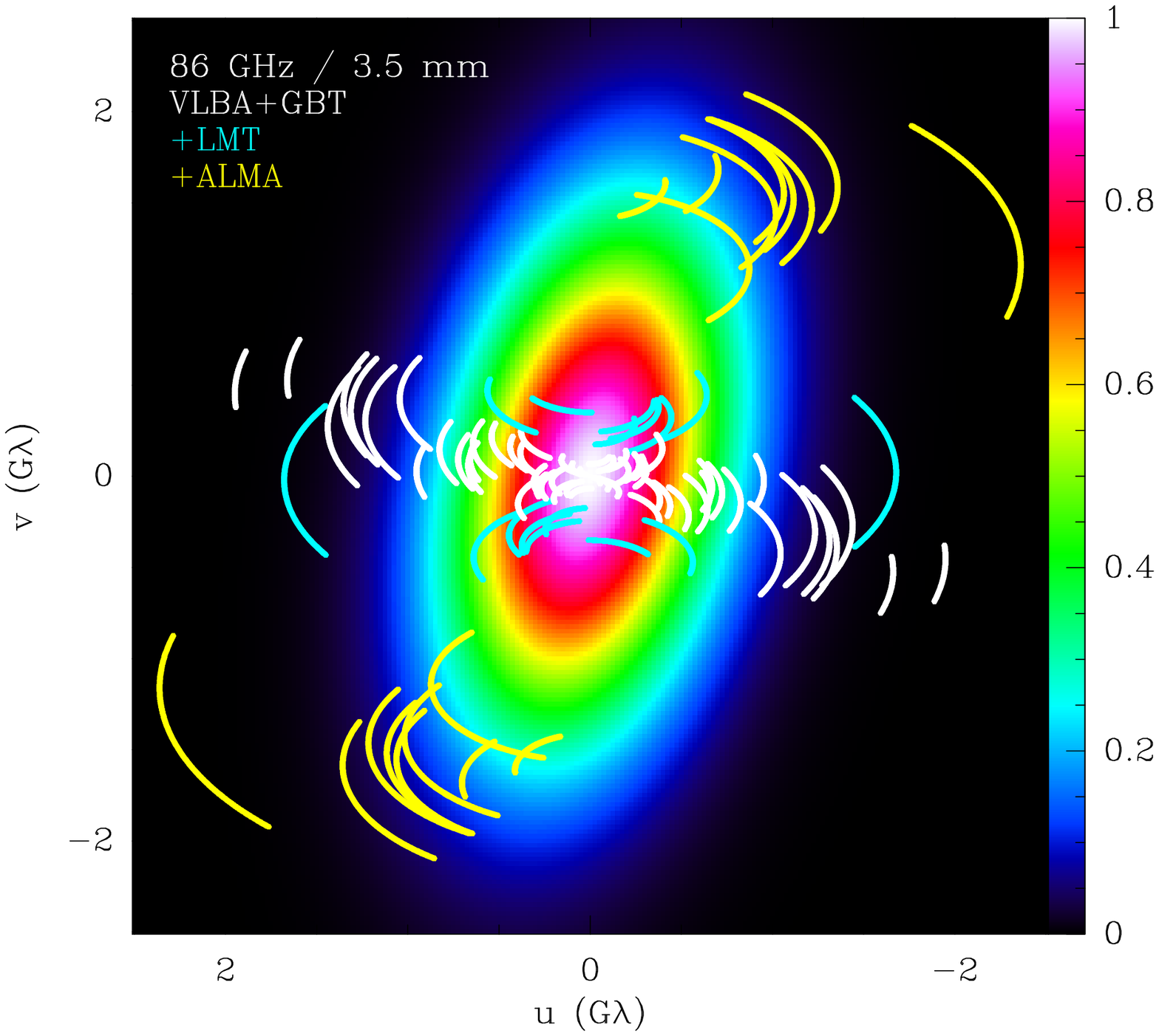}}
\caption{Plot of the elliptical Gaussian scattering kernel
  $\tilde{G}(\mathbf{u})$ in the $(u,v)$-plane at 1.3~mm (\emph{top})
  and 3.5~mm (\emph{bottom}), with baseline tracks overplotted.  White
  tracks show the $(u,v)$ coverage attainable with the EHT (including
  the LMT and ALMA) at 1.3~mm and the VLBA plus GBT at 3.5~mm.  The
  addition of the LMT (cyan) to the 3.5~mm observing array provides
  many baseline tracks that are not heavily scattered.  Baselines from
  ALMA (yellow) to the LMT, GBT, and some VLBA stations may also have
  detectable flux.
\label{uv-gaussian}
}
\end{figure}

Because the elliptical scattering kernel is strictly positive, its
effects are invertible.  Measured visibilities can be divided by
$\tilde{G}(\mathbf{u})$ to recover estimates of the visibilities of
the \emph{unscattered} image.  Of course, the loss in the
signal-to-noise ratio, $S/N \equiv
\tilde{I}(\mathbf{u})/\tilde{N}(\mathbf{u})$, where
$\tilde{N}(\mathbf{u})$ represents the noise of the measured
visibility, cannot be recovered, since the interferometer senses the
scattered image.  This places a natural limit on the applicability of
the inversion technique.  At very large $\mathbf{u}$, $S/N \rightarrow
0$ and $\tilde{G}(\mathbf{u}) \rightarrow 0$, with the result that
division by $\tilde{G}(\mathbf{u})$ amplifies the noise (although such
data points would have very little weight in most image reconstruction
algorithms due to their low $S/N$).  This limit is not applicable to
ground-based VLBI of Sgr~A* at 1.3~mm, where $\tilde{G}(\mathbf{u}) >
0.19$ over the entire range of $\mathbf{u}$ that will be covered by
EHT baselines (Figure~\ref{uv-gaussian}).

An alternate strategy is to first reconstruct the scattered image and
then attempt deconvolution.  To obtain the unscattered image we must
solve an inverse problem that deconvolves the known scattering
$G(\mathbf{x})$ from the scattered image.  This inverse problem,
termed non-blind deconvolution, has been studied extensively in signal
and image processing.  Without noise the solution to the inverse
problem is trivial: Fourier transform the image, divide by
$\tilde{G}(\mathbf{u})$, and inverse transform back to the image
domain.  However, in practice, noise dominates the high spatial
frequencies of a reconstructed image.  Dividing by
$\tilde{G}(\mathbf{u})$ strongly amplifies this noise, introducing
potentially severe artifacts into the image.  Many different non-blind
deconvolution approaches exist that vary greatly in their speed and
sophistication in order to address this problem
\citep[e.g.,][]{krishnan2009,zoran2011,joshi2009}, but the Wiener
deconvolution filter is perhaps the simplest and most general
deconvolution approach \citep{wiener1949}.  The Wiener filter,
\begin{equation}
\tilde{G}^{-1}_\mathrm{W}(\mathbf{u}) =
\frac{1}{\tilde{G}({\mathbf{u}})} \left[
  \frac{|\tilde{G}(\mathbf{u})|^2} {|\tilde{G}(\mathbf{u})|^2 +
        K} \right],
\end{equation}
can be defined in terms of a tunable parameter $K$ \citep{russ2011}.
As $K \rightarrow 0$, $\tilde{G}^{-1}_\mathrm{W}(\mathbf{u})
\rightarrow \tilde{G}^{-1}(\mathbf{u})$, resulting in a sharper image
but potentially unbounded amplification of high-frequency noise.  In
theory, the optimal value of $K$ is inversely proportional to the
square of $S/N$, but the noise in the Fourier domain is neither
constant nor known a priori.  Indeed, the noises of the visibilities
from which the image is reconstructed are typically unequal due to
differing telescope sensitivities, and the image reconstruction
process (which must necessarily fill in information from unmeasured
Fourier components) introduces additional noise in the Fourier domain.
In practice, $K$ is treated as a tunable parameter\footnote{When an
  image is reconstructed from corrected visibilities, the $S/N$ of
  each data point is retained, so highly amplified low-$S/N$ data do
  not corrupt the reconstructed image.  In contrast, when an image is
  reconstructed from the uncorrected visibilities and deconvolution is
  attempted in the image domain, the $S/N$ information associated with
  each Fourier component is not preserved.}.

\section{Methods} \label{methods}

We demonstrate how scattering may be mitigated in practice by
simulating observations of Sgr~A* at 1.3~mm.  We include the effects
of scattering, generate synthetic data, correct the visibility
amplitudes, reconstruct the image, and assess image fidelity relative
to the model image.

\subsection{Data Simulation}

Our input image is a semi-analytic radiatively inefficient accretion
flow model of Sgr~A* using the best-fit model parameters from
\citet{broderick2011}.  This image was scattered using the parameters
given by \citet{bower2006}.  Simulated 1.3~mm EHT data were produced
using the MIT Array Performance Simulator
(MAPS).\footnote{\url{http://www.haystack.mit.edu/ast/arrays/maps/index.html}}
Telescopes in the simulated array included the Submillimeter Array and
James Clerk Maxwell Telescope on Mauna Kea, the Arizona Radio
Observatory Submillimeter Telescope, the Combined Array for Research
in Millimeter-wave Astronomy, the Large Millimeter Telescope (LMT),
the Atacama Large Millimeter/submillimeter Array (ALMA), the Institut
de Radioastronomie Millim\'{e}trique (IRAM) 30-m telescope on Pico
Veleta, the IRAM Plateau de Bure Interferometer, and the South Pole
Telescope.  Further details of the simulated observing array can be
found in \citet{lu2014}.

Our simulations include realistic thermal noise but neglect the effect
of calibration errors on visibility amplitudes.  It is difficult to
estimate what the magnitude of calibration errors will be in the era
when the EHT has enough sensitivity and baseline coverage to image
Sgr~A*.  In contrast with the anticipated capabilities of the EHT in
the next few years, previously published EHT observations of Sgr~A*
\citep{doeleman2008,fish2011} are based on comparatively low-S/N data
taken with an array consisting of only three VLBI sites.  Accurate
amplitude calibration has been challenging for a variety of reasons,
such as the limited sensitivity of the array and the paucity of
redundancy in the data.  The EHT of the near future will almost
certainly do better through a combination of higher sensitivity,
enabled by much wider bandwidths; scheduling designed to improve a
priori data calibration, itself enabled by higher sensitivity;
calibration constraints provided by amplitude closure, which requires
at least four VLBI stations; imaging algorithms which treat closure
amplitudes as fundamental observables that are inherently robust
against amplitude calibration errors; use of prior phase information
to provide partial phase calibration of the array; and optimized data
processing.  Initial images of the quiescent millimeter-wavelength
emission from Sgr~A* are likely to be of lower fidelity than presented
in this work for a number of reasons, including calibration errors,
refractive phase noise on long baselines
(Section~\ref{refractive-noise}), and rapid source variability
(Section~\ref{variability}).  Nevertheless, it is instructive to
examine scattering mitigation on ideal simulated data in order to
explore the possibilities and limitations of mitigation techniques.

\subsection{Image Reconstruction}

The simulated data were imaged using the BiSpectrum Maximum Entropy
Method \citep[BSMEM;][]{buscher1994}, as described in detail in
\citet{lu2014}.  Developed by the infrared and optical interferometry
community, BSMEM differs from classical centimeter-wavelength
interferometric imaging techniques in two key ways that are well
suited to millimeter-wavelength VLBI imaging.  First, variations in
the tropospheric delay impose phase fluctuations that are too rapid to
calibrate out using standard phase-referencing techniques.  Since the
atmospheric phase contributions are antenna-based, their contributions
cancel along a closed loop of antennas
\citep[e.g.,][]{jennison1958,rogers1974}.  BSMEM thus treats closure
phases on triangles of stations, rather than visibility phases on
baselines, as the fundamental phase observables for imaging.  Second,
the early EHT, like optical interferometers, will have sparse $(u,v)$
coverage.  Deconvolution-based imaging techniques such as CLEAN
\citep{hogbom1974} and Multi-Scale CLEAN
\citep{cornwell2008,greisen2009} reconstruct an image by Fourier
inverting the sampled visibilities to produce the so-called dirty map
and then deconvolving the dirty beam (also known as the point spread
function) in the image domain.  These techniques work well when
imaging simple structures with arrays consisting of a large number of
antennas (e.g., the Jansky Very Large Array).  However, when the array
consists of very few antennas, there are large unsampled areas in the
$(u,v)$ plane, producing a dirty beam that has large sidelobes.  In
any case, the dirty beam is not positive-definite.  Small errors
during deconvolution can thus produce large artifacts in the image
domain, severely limiting image fidelity.  In contrast, forward
imaging methods such as BSMEM avoid deconvolution by finding best-fit
images that are directly consistent with the observables, using a
priori knowledge and suitable regularizers.  Unlike with CLEAN, images
reconstructed with BSMEM are not normally convolved with a restoring
beam, which in this case would significantly degrade the resolution of
the reconstructed images.

\subsection{Image Fidelity Analysis}

We assessed image quality using both pixel-based and feature-based
metrics.  In the first category, the mean square error (MSE)
quantifies the mean square pixel-by-pixel intensity difference between
a truth image and a reconstructed image, normalized by the sum of the
squares of the pixel intensities in the reference image.  The MSE is
equivalent to the metric used to assess image quality in the biennial
interferometric imaging beauty contest \citep[][most
  recently]{baron2012}.  In the second category, motivated by human
visual perception, the structural dissimilarity (DSSIM) index, derived
from the structural similarity (SSIM) index \citep{wang2004,loza2009},
quantifies differences in luminance, contrast, and structure between
two images.  Formal definitions of the MSE and DSSIM metrics are given
in \citet{lu2014}.  For both indices, lower values correspond to
better image quality.  Because absolute positional information is lost
when reconstructing images using closure phase information, the images
are cross-registered to the reference image at sub-pixel accuracy
before calculating these metrics.

As an additional measure of artifacts introduced by Wiener
deconvolution, we will also quote the ratio of the largest positive
pixel to the absolute value of the largest negative pixel (max/min).
This metric is not directly applicable to images reconstructed with
BSMEM, which are constrained to be positive-definite.

\section{Results} \label{results}

We seek to answer three questions.  First, does using corrected
visibility amplitudes mitigate the effects of scatter broadening?
Second, do corrected visibilities produce a higher-fidelity image than
Wiener deconvolution of the scatter-broadened image?  Third, are
uncertainties in the scattering parameters at 1.3~mm small enough to
permit scattering mitigation by these methods?

\begin{figure*}
\begin{center}
\resizebox{!}{0.24\hsize}{\includegraphics{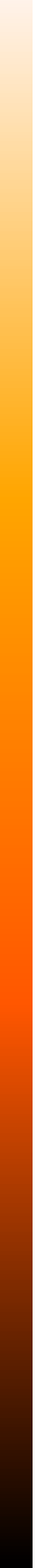}}
\resizebox{0.24\hsize}{!}{\includegraphics{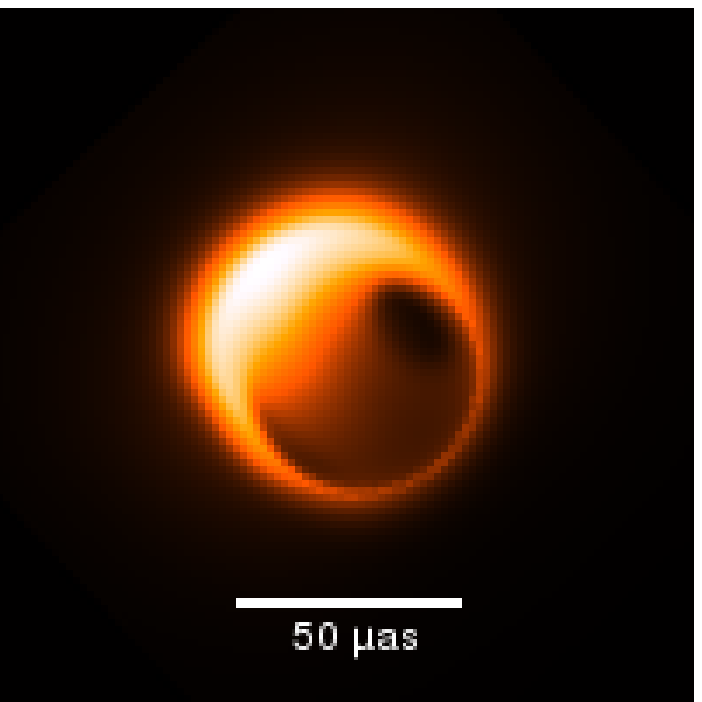}}
\resizebox{0.24\hsize}{!}{\includegraphics{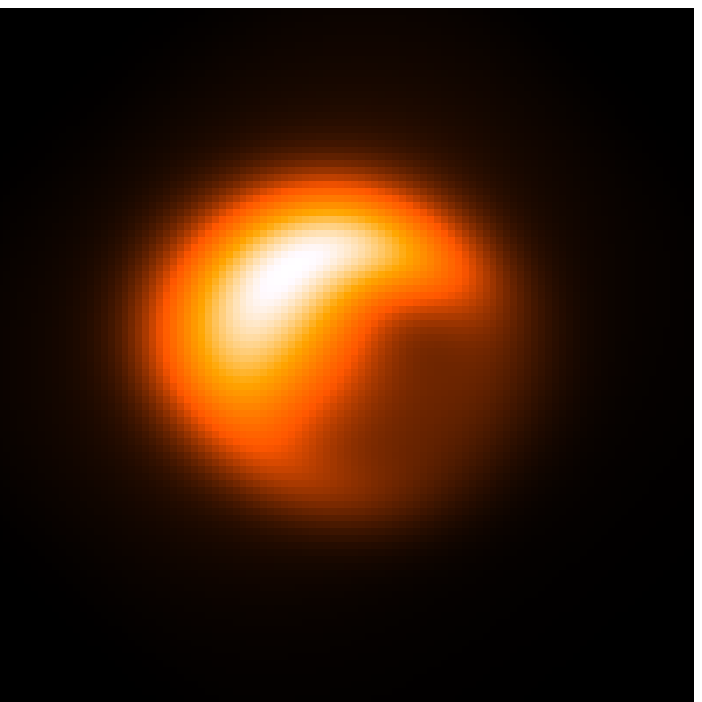}}
\resizebox{0.24\hsize}{!}{\includegraphics{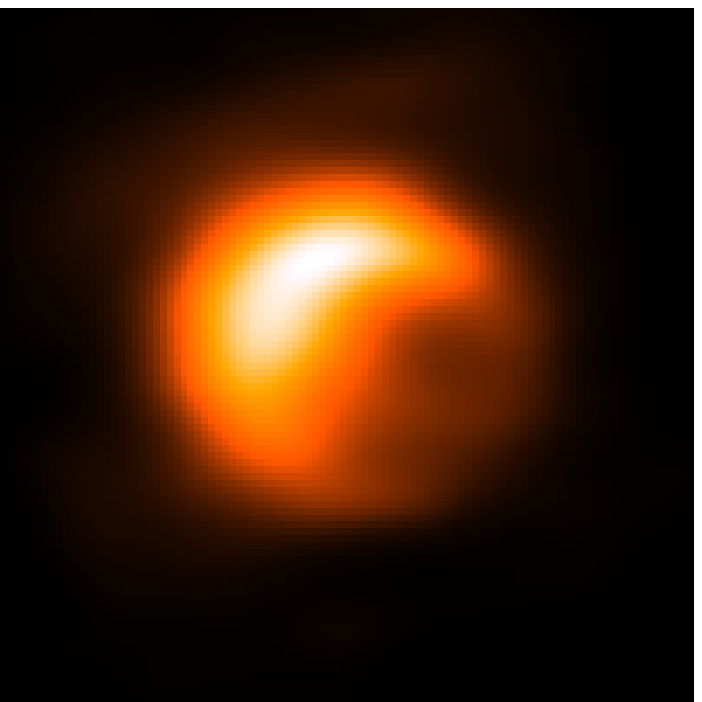}}
\resizebox{0.24\hsize}{!}{\includegraphics{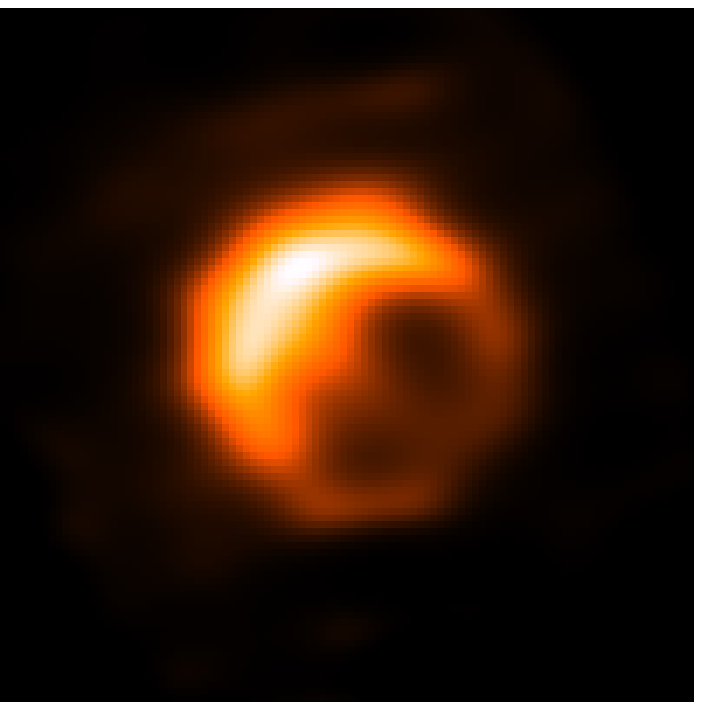}}
\end{center}
\caption{Imaging simulation of Sgr~A* with the EHT.  The model image
  (\emph{first panel}) for a semi-analytic accretion flow around
  Sgr~A* \citep{broderick2011} is convolved with the known scattering
  kernel (\emph{second}; MSE 0.087 and DSSIM 0.117 with first panel as
  the reference image).  The BSMEM reconstruction of the scattered
  image (\emph{third}; MSE 0.099, DSSIM 0.154) can be improved upon by
  dividing each synthetic visibility by the Fourier transform of the
  scattering kernel; the subsequent BSMEM reconstruction produces an
  image (\emph{fourth}; MSE 0.024, DSSIM 0.076) much closer to the
  unscattered original.  The image reconstruction using the corrected
  visibilities is able to clearly detect the shadow and photon ring
  associated with the black hole.  A linear transfer function
  (\emph{far left}) is used in each panel in this and subsequent
  figures.
\label{reconstructions}
}
\end{figure*}

Correcting visibility amplitudes does indeed partially mitigate the
effects of scatter broadening.  The reconstructed image from the
corrected visibilities is much sharper than would be recovered from
the uncorrected visibilities (Figure~\ref{reconstructions}), as
confirmed by the MSE (0.099 uncorrected, 0.024 corrected) and DSSIM
(0.154 uncorrected, 0.076 corrected) values.  The image reconstructed
from the corrected visibilities is able to recover the important
physical features introduced by GR, including the dark shadow of the
black hole and the associated bright photon ring at its edge
\citep{broderick2006,dexter2009,johannsen2010a,johannsen2010b}.
Although these features are present in the reconstruction produced
from the uncorrected visibilities, they are much less prominent,
reflecting the fact that they are also less prominent in the scattered
model image itself.

\begin{figure*}
\begin{center}
\resizebox{0.24\hsize}{!}{\includegraphics{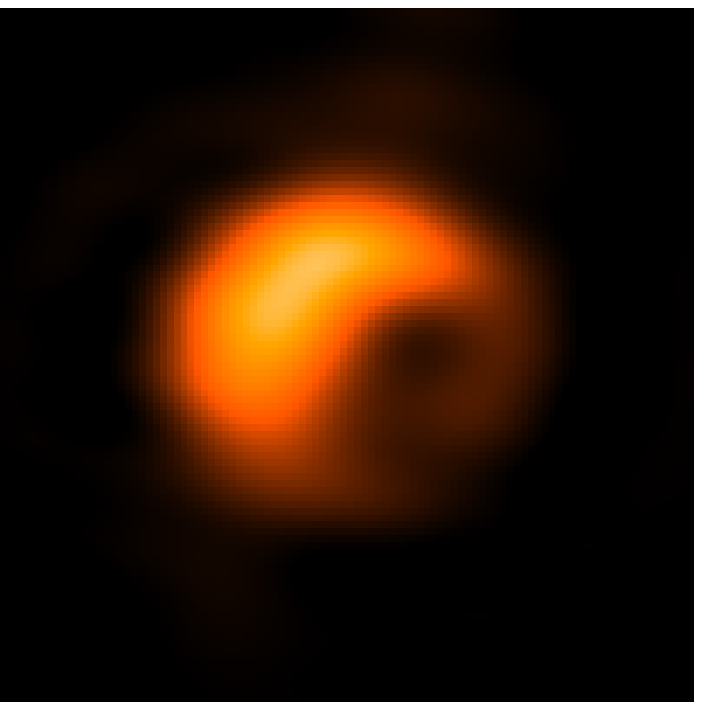}}
\resizebox{0.24\hsize}{!}{\includegraphics{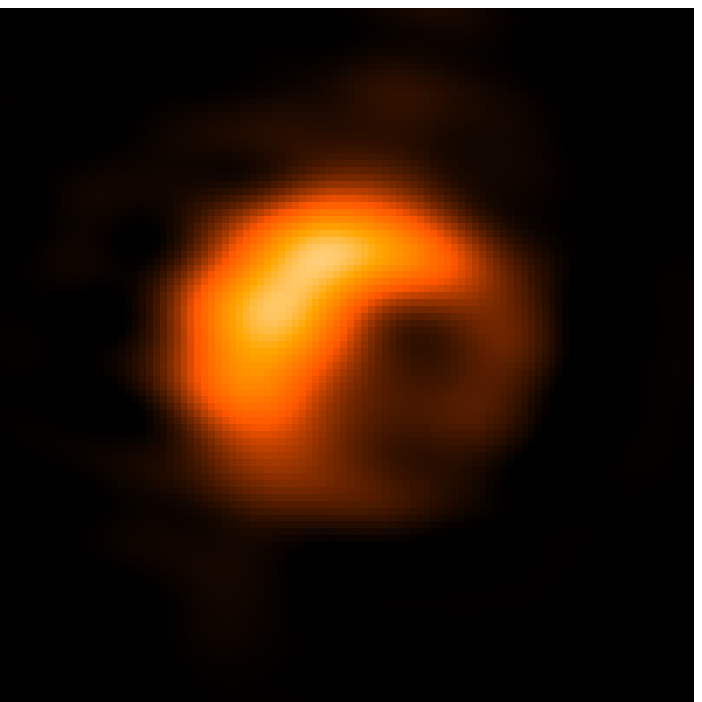}}
\resizebox{0.24\hsize}{!}{\includegraphics{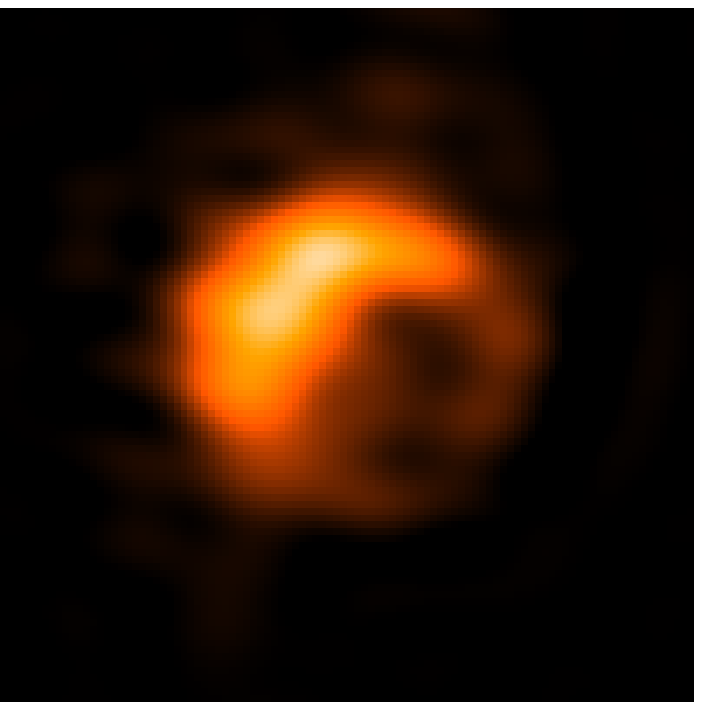}}
\resizebox{0.24\hsize}{!}{\includegraphics{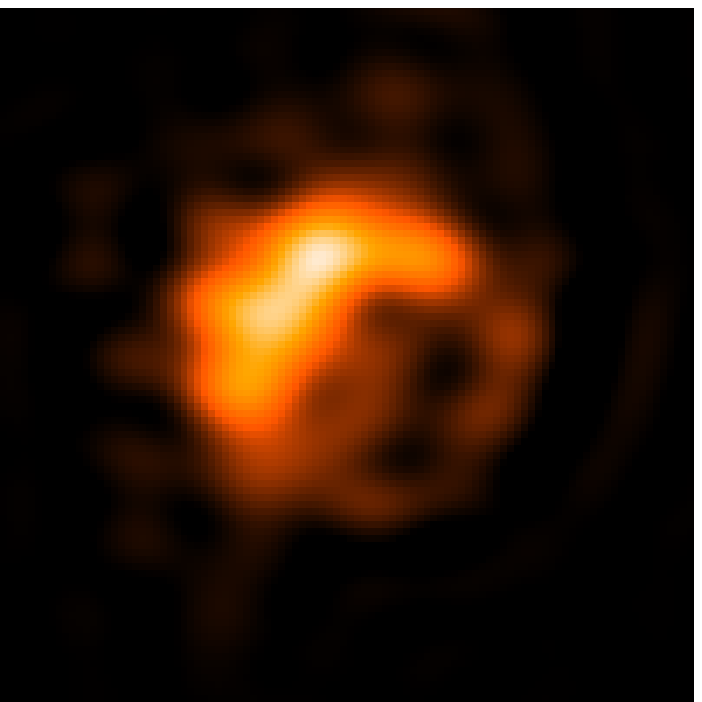}}
\end{center}
\caption{Wiener deconvolution of the scattering kernel from the
  scattered image reconstruction (third panel of
  Figure~\ref{reconstructions}).  As the noise threshold is lowered
  (by factors of $\sim 3$ between panels from left to right), the
  deconvolved image becomes sharper, but high-frequency noise
  increasingly dominates (left to right: max/min 36.6, 26.1, 19.3,
  13.7).  The image produced by correcting amplitudes in the
  visibility domain (fourth panel of Figure~\ref{reconstructions}) has
  higher fidelity than any attempt to deconvolve the scattering in
  images produced from the uncorrected amplitudes (left to right: MSE
  0.058, 0.046, 0.040, 0.045; DSSIM 0.148, 0.140, 0.140, 0.160).  An
  identical linear transfer function is used in all panels.
\label{wiener}
}
\end{figure*}

Correcting the visibilities before imaging fares better than
reconstructing the scattered image and attempting deconvolution in the
image domain (Figure~\ref{wiener}).  As the Wiener deconvolution
  parameter $K$ is lowered, features in the deconvolved image become
  sharper, but the strength of artifacts introduced by deconvolution
  become enhanced.  A key distinction between these scattering
  mitigation methods is that the image directly reconstructed from
  corrected visibilities is positive-definite, while the deconvolved
  images have negative-pixel artifacts.

\begin{figure*}
\begin{center}
\resizebox{0.24\hsize}{!}{\includegraphics{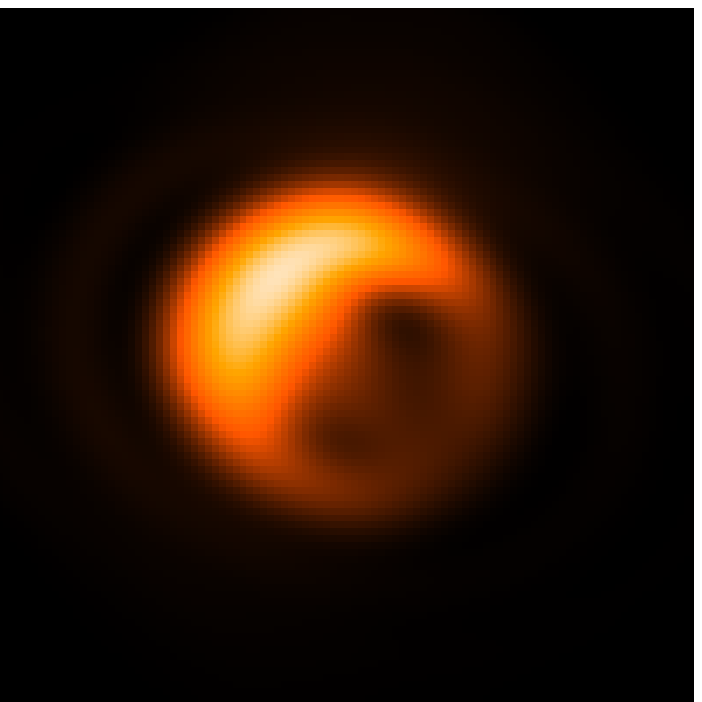}}
\resizebox{0.24\hsize}{!}{\includegraphics{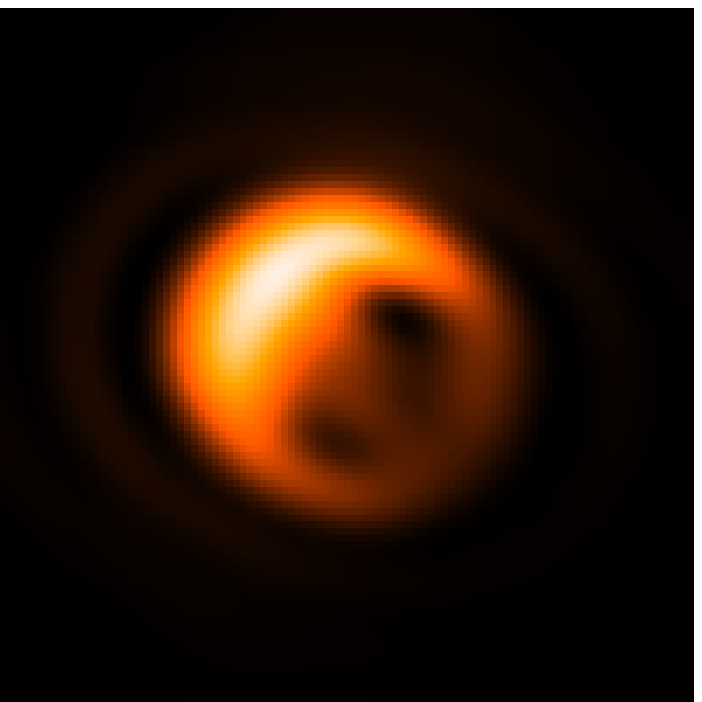}}
\resizebox{0.24\hsize}{!}{\includegraphics{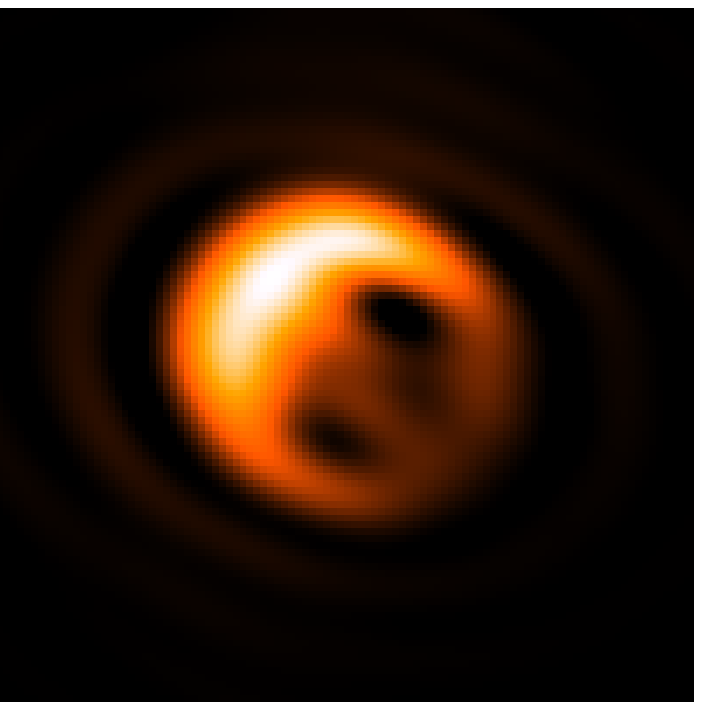}}
\end{center}
\caption{The model image convolved with one scattering kernel and
  Wiener deconvolved with another (\emph{left to right}: the large
  kernel unscattered with the small kernel, fiducial unscattered with
  fiducial, and small unscattered with large, as described in
  Section~\ref{results}).  The close similarity in all cases (left to
  right: MSE 0.026, 0.017, 0.021; DSSIM 0.046, 0.054, 0.093)
  demonstrates that minor uncertainties in the scattering kernel are
  unlikely to pose major problems at 1.3~mm, although image artifacts
  become significantly more prominent if the image is overcorrected
  with Wiener deconvolution (left to right: max/min 533.6, 304.7,
  18.4).
\label{scattering-uncertainties}
}
\end{figure*}

To determine whether uncertainties in the properties of the scattering
are large enough to hamper mitigation, we considered three scattering
kernels at 1.3~mm: a fiducial kernel from \citet{bower2006} ($1.309
\times 0.64$~mas\,cm$^{-2}$ at $78\degr$ east of north), a small
kernel from the $-1\,\sigma$ errors ($1.294 \times
0.59$~mas\,cm$^{-2}$ at $77\degr$), and a large kernel from the
$+1\,\sigma$ errors of \citet{shen2005} ($1.41 \times
0.75$~mas\,cm$^{-2}$ at $80\degr$).  The source image was convolved
and Wiener deconvolved with different combinations of these kernels
(Figure~\ref{scattering-uncertainties}).  In order to isolate and
clearly illustrate the effects of mitigating scattering with incorrect
parameters, convolution and deconvolution were applied to the model
image itself rather than the BSMEM reconstruction thereof.
Under-removing scattering results in a slightly more blurred image
than ideal, but the result is nevertheless a substantial improvement
upon the scattered model image.  Over-removing scattering results in a
sharper image of Sgr~A* at the expense of substantially larger image
artifacts.  Nevertheless, all cases significantly improve upon the
resulting image as measured by both the MSE and DSSIM metrics.

\section{Discussion}

\subsection{Validity of the Scattering Approximation} \label{refractive-noise}

We have demonstrated that the effects of idealized interstellar
scattering of Sgr~A* at 1.3~mm can be partially mitigated by
correcting visibilities before imaging.  However, there are two
potential ways in which real scattering may deviate from this
approximation on long baselines: the phase structure function may
deviate from an $\alpha \approx 2$ power law, and the scattering
kernel may introduce refractive noise.  We now estimate the influence
of each of these uncertainties.

The phase structure function can be better characterized by continued
observations at longer wavelengths and on nearby targets.  As
previously discussed, the scattering response must be an anisotropic
Gaussian on short baselines.  On longer baselines, the most plausible
transition is to a Kolmogorov spectrum, which predicts that the image
will be less scattered (and therefore that scattering mitigation will
be less important) than would be predicted by the Gaussian regime.
Because the scattering for Sgr A* at 1.3 mm may have $r_0 \sim
r_\mathrm{in}$, improved understanding of the scattering will be vital
for baselines ${\gtrsim}3$~G$\lambda$.  The recent discovery of
substructure in the scattered image of Sgr~A* at 1.3~cm wavelength,
which supports the Kolmogorov transition at wavelengths near a
millimeter, provides one promising avenue to better characterize the
scattering \citep{gwinn2014}.  Deeper studies of similarly scattered
objects, such as the Galactic center magnetar SGR J1745$-$29, may also
provide guidance \citep{bower2014}.  In any case, the amplitude
portion of the scattering kernel $\tilde{G}(\mathbf{u})$ will be a
predictable function of $\alpha$, $r_0$, and $r_\mathrm{in}$
\citep[eq.~(3.1.1) of][]{goodman1989}.  Fine adjustments of the
scattering kernel, and thus the unscattered image, can be achieved
through blind-deconvolution methods \citep{levin2009,levin2011}.

We can also estimate the level of refractive noise for Sgr~A*.  The
$\lambda^2$ scaling of the scattering at longer wavelengths suggests
that the inner scale of the turbulence is larger than the phase
coherence length on the scattering screen.  Taking $\alpha = 5/3$ in
the turbulent regime, the rms refractive noise
$\sigma_\mathrm{R}(\mathbf{u}) \equiv \left \langle \sqrt{\left|
  \Delta \tilde{I}(\mathbf{u}) \right|^2} \right \rangle$, normalized
by the zero baseline flux, for a point source is
\begin{equation}
\sigma_\mathrm{R}(\mathbf{u}) \approx 0.37 \left( \frac{r_0}{r_\mathrm{F}}
\right)^{1/3} \left( \frac{ |\mathbf{u}|\lambda }{ (1 + M) r_0 } \right)^{-5/6} \left(
\frac{r_\mathrm{in}}{r_0} \right)^{1/6}
\end{equation}
\citep[eq.~(5.1.2) of][]{goodman1989}.
At 1.3~mm, the Fresnel scale is
\begin{equation}
r_\mathrm{F} = \sqrt{ \frac{d}{k} \frac{M}{(1+M)^2} } \approx 10^5~\mathrm{km}
\end{equation}
for $M \sim 1$ \citep{bower2014}. The inner scale is most likely
hundreds of kilometers \citep[see][]{spangler1990}, especially in
light of recent measurements that disfavor a unique scattering
environment \citep{bower2014}.  This replacement anticipates
$\sigma_\mathrm{R}(\mathbf{u}) \approx 0.15$ on a 3~G$\lambda$
east-west baseline, for instance.

The refractive noise is correlated among different baselines, with a
decorrelation length comparable to the diffractive scale
\citep{goodman1989}.  For a particular realization of the refractive
noise, it will be correlated on all EHT baselines
(Figure~\ref{uv-gaussian}).  Thus, by normalizing measurements by the
zero-baseline flux, the noise on long baselines will be partially
suppressed.

Refractive noise will additionally be suppressed by the finite
(non-pointlike) extent of Sgr~A*.  Thus, even on long baselines
(${\gtrsim}3$~G$\lambda$), the refractive phase ``jitter'' may be a
$10^\circ$ or less with a characteristic timescale of variation of
about 1 day (the refractive timescale).  This phase would be constant
over the duration of an observing night and across the entire
observing bandwidth.  The most sensitive evidence of these effects may
be zero-mean variations in closure phase on timescales that are
compatible with the refractive timescale (but see
Section~\ref{variability}).  Larger variations could indicate atypical
properties of the turbulence that gives rise to the scattering along
the line of sight to the Galactic center.

Refractive phase noise may introduce a small jitter in the apparent
position of Sgr~A*.  This may introduce additional noise into
measurements of the position wander of Sgr~A* on interday timescales
but will be constant on shorter timescales corresponding to orbital
periods in the accretion flow.  It will also be negligible when
averaged over the longer timescales necessary for measuring the
parallax and proper motion of Sgr~A* \citep{reid2008,broderick2011b}.

Since refractive phase noise will have zero mean, a strategy to
mitigate the effects of refractive phase noise is to observe Sgr~A*
for multiple nights.  Phase quantities measured at identical points in
the $(u,v)$ plane can be averaged across all observing nights to
produce a single dataset.  As the particular realizations of
refractive phase noises will be different from day to day, averaging
many days of data together effectively produces a dataset that is
sensitive to the idealized (real) ensemble-average image scattering
regime.  As the number of nights of data increases, the scattering
kernel will asymptote to the real, invertible ensemble-average kernel.

\subsection{Connections to Variability}\label{variability}

Refractive noise will introduce small distorions into the observed
image of Sgr~A*.  However, these distortions will be small compared to
intrinsic changes in the emitting material around the black hole.
Sgr~A* exhibits variability and flaring activity across the spectrum
\citep[e.g.,][]{eckart2006}.  At millimeter wavelengths, this is
likely accompanied by structural changes or the appearance of hot
spots in the accretion flow.  GR magnetohydrodynamic simulations of
the dynamic accretion flow
\citep[e.g.,][]{noble2007,moscibrodzka2009,dexter2010} demonstrate
that inhomogeneities can imprint large changes upon measured closure
phases, as can nonthermal hot-spots
\citep{broderick2005,doeleman2009}.  These variations will be evident
on timescales of minutes.  The innermost stable circular orbital
period of Sgr~A* is about half an hour if the black hole spin is close
to zero and shorter still if the black hole is spinning rapidly, and
the light-crossing time for a region that is one gravitational radius
across is $GMc^{-3} \approx 20$~s.

Standard aperture-synthesis techniques rely on Earth rotation to
change the projected baselines over the course of many hours,
providing greater coverage in the $(u,v)$ plane and hence higher
imaging fidelity.  However, visibilities obtained at different times
will correspond to different images.  Naive application of the
Earth-rotation aperture synthesis technique may fail because there
will not be a single image that is consistent with all measured
visibilities in the $(u,v)$ plane.  To obtain an image of the average
quiescent emission around an intrinsically variable Sgr~A*, it may
therefore be necessary to average visibilities across multiple nights
(Lu et al., in preparation), although non-imaging techniques can still
recover detailed spatial information on much shorter timescales
\citep{doeleman2009,fish2009}.  The effects of refractive phase noise,
small compared to those of source variability, will also be suppressed
by this averaging process (Section~\ref{refractive-noise}).

\subsection{Applicability at 3.5~mm}

The technique described in this Letter will be applicable to Sgr~A* at
other wavelengths.  For Sgr~A* at 1.3~mm, the longest EHT baselines
exceed the diffractive scale by less than a factor of two
(Figure~\ref{uv-gaussian}).  At 3.5~mm, most baselines among the inner
7 telescopes of the Very Long Baseline Array (VLBA) are shorter than
the diffractive scale.  Baselines to the Large Millimeter Telescope
and Green Bank Telescope may nevertheless produce detectable fringes,
since their large collecting area will offset additional losses from
scattering on longer baselines.  Scattering should not be so strong on
baselines between the LMT and VLBA stations (with the exception of
Mauna Kea) to prevent detection of Sgr~A* (Figure~\ref{uv-gaussian}).
Baselines between phased ALMA and continental North America, though
long, have a favorable orientation relative to the scattering ellipse.
It is possible that Sgr~A* may be detected on the very sensitive
ALMA-LMT and ALMA-GBT baselines.  However, the amount of correlated
flux density on long baselines to ALMA (with projected baseline
lengths ranging from $\sim$ 5000 to 7200~km between ALMA and the VLBA
sites of Fort Davis, Kitt Peak, Los Alamos, North Liberty, and Pie
Town) will be strongly dependent on the intrinsic size of Sgr~A* at
3.5~mm in these directions, estimated to be $\gtrsim
100~\mu$as\ \citep{bower2006,lu2011}.

\section{Conclusions}

We have demonstrated that the scatter broadening of Sgr~A* at 1.3~mm
can be significantly mitigated.  The predominant effect of
interstellar scattering is to decrease the measured visibility
amplitudes, especially on long baselines.  As long as the observing
array is sufficiently sensitive to obtain fringes on these baselines,
visibility amplitudes can be corrected for this effect, recovering
most of the unscattered structure of Sgr~A*.  Reconstructing the image
from corrected visibilities produces a higher-fidelity image than
attempting deconvolution after reconstructing an image from the
uncorrected visibilities.

This result is of direct relevance for imaging Sgr~A* with the EHT at
1.3~mm and may be applicable to longer wavelengths as well.  It is
also possible to construct non-imaging VLBI observables, such as
closure phase and fractional polarization, that are unbiased by
scatter broadening.

\acknowledgments

The Event Horizon Telescope is supported by grants from the National
Science Foundation (NSF) and from the Gordon and Betty Moore
Foundation (\#GBMF-3561).
K.L.B.~is partially supported by an NSF Graduate Fellowship.
D.P.~is supported by NASA/NSF TCAN award NNX14AB48G and by NSF award
AST 1312034.
R.N.~was supported in part by NSF grant AST 1312651.
A.E.B.~receives financial support from the Perimeter Institute for
Theoretical Physics and the Natural Sciences and Engineering Research
Council of Canada through a Discovery Grant.  Research at Perimeter
Institute is supported by the Government of Canada through Industry
Canada and by the Province of Ontario through the Ministry of Research
and Innovation.

\


\begin{thebibliography}{}

\bibitem[Akiyama et al.(2013)]{akiyama2013} Akiyama, K., Takahashi,
  R., Honma, M., Oyama, T., \& Kobayashi, H.\ 2013, \pasj, 65, 91

\bibitem[Armstrong et al.(1995)]{armstrong1995} Armstrong, J.~W.,
  Rickett, B.~J., \& Spangler, S.~R.\ 1995, \apj, 443, 209

\bibitem[Bardeen(1973)]{bardeen1973} Bardeen, J.~M.\ 1973, in Black
  Holes (Les Astres Occlus), ed.\ C.\ DeWitt \& B.~S.~DeWitt (New
  York: Gordon \& Breach), 215

\bibitem[Baron et al.(2012)]{baron2012} Baron, F., et al.\ 2012,
  \procspie, 8445

\bibitem[Bower et al.(2004)]{bower2004} Bower, G.~C., Falcke, H.,
  Herrnstein, R.~M., Zhao, J.-H., Goss, W.~M., \& Backer, D.~C.\ 2004,
  Science, 304, 704

\bibitem[Bower et al.(2006)]{bower2006} Bower, G.~C., Goss, W.~M.,
  Falcke, H., Backer, D.~C., \& Lithwick, Y.\ 2006, \apjl, 648, L127

\bibitem[Bower et al.(2014)]{bower2014} Bower, G.~C., et al.\ 2014,
  \apjl, 780, L2

\bibitem[Broderick et al.(2011a)]{broderick2011} Broderick, A.~E.,
  Fish, V.~L., Doeleman, S.~S., \& Loeb, A.\ 2011a, \apj, 735, 110

\bibitem[Broderick \& Loeb(2005)]{broderick2005} Broderick, A.~E., \&
  Loeb, A.\ 2005, \mnras, 363, 353

\bibitem[Broderick \& Loeb(2006)]{broderick2006} Broderick, A.~E., \&
  Loeb, A.\ 2006, \mnras, 367, 905

\bibitem[Broderick et al.(2011b)]{broderick2011b} Broderick, A.~E.,
  Loeb, A., \& Reid, M.~J.\ 2011b, \apj, 735, 57

\bibitem[Buscher(1994)]{buscher1994} Buscher, D.~F.\ 1994, Very High
  Angular Resolution Imaging, 158, 91

\bibitem[Cornwell(2008)]{cornwell2008} Cornwell, T.~J.\ 2008, IEEE
  Journal of Selected Topics in Signal Processing, 2, 793

\bibitem[Dexter et al.(2009)]{dexter2009} Dexter, J., Agol, E., \&
  Fragile, P.~C.\ 2009, \apjl, 703, L142

\bibitem[Dexter et al.(2010)]{dexter2010} Dexter, J., Agol, E.,
  Fragile, P.~C., \& McKinney, J.~C.\ 2010, ApJ, 717, 1092

\bibitem[Doeleman et al.(2009b)]{whitepaper} Doeleman, S., et
  al.\ 2009b, astro2010: The Astronomy and Astrophysics Decadal
  Survey, 2010, 68

\bibitem[Doeleman et al.(2009a)]{doeleman2009} Doeleman, S.~S., Fish,
  V.~L., Broderick, A.~E., Loeb, A., \& Rogers, A.~E.~E.\ 2009a, \apj,
  695, 59

\bibitem[Doeleman et al.(2001)]{doeleman2001} Doeleman, S.~S., et
  al.\ 2001, \aj, 121, 2610

\bibitem[Doeleman et al.(2008)]{doeleman2008} Doeleman, S.~S., et
  al.\ 2008, \nat, 455, 78

\bibitem[Eckart et al.(2006)]{eckart2006} Eckart, A., et al.\ 2006,
  \aap, 450, 535

\bibitem[Falcke et al.(2000)]{falcke2000} Falcke, H., Melia, F., \&
  Agol, E.\ 2000, \apjl, 528, L13

\bibitem[Falcke et al.(2009)]{falcke2009} Falcke, H., Markoff, S., \&
  Bower, G.~C.\ 2009, \aap, 496, 77

\bibitem[Fish et al.(2009)]{fish2009} Fish, V.~L., Doeleman, S.~S.,
  Broderick, A.~E., Loeb, A., \& Rogers, A.~E.~E.\ 2009, \apj, 706,
  1353

\bibitem[Fish et al.(2011)]{fish2011} Fish, V.~L., et al.\ 2011,
  \apjl, 727, L36

\bibitem[Fish et al.(2013)]{fish2013} Fish, V.~L., et al.\ 2013,
  High-Angular-Resolution and High-Sensitivity Science Enabled by
  Beamformed ALMA, arXiv:1309.3519

\bibitem[Ghez et al.(2008)]{ghez2008} Ghez, A.~M., et al.\ 2008, \apj,
  689, 1044

\bibitem[Gillessen et al.(2009b)]{gillessen2009b} Gillessen, S.,
  Eisenhauer, F., Fritz, T.~K., Bartko, H., Dodds-Eden, K., Pfuhl, O.,
  Ott, T., \& Genzel, R.\ 2009b, \apjl, 707, L114

\bibitem[Gillessen et al.(2009a)]{gillessen2009a} Gillessen, S.,
  Eisenhauer, F., Trippe, S., Alexander, T., Genzel, R., Martins, F.,
  \& Ott, T.\ 2009a, \apj, 692, 1075

\bibitem[Goldreich \& Sridhar(1995)]{goldreich1995} Goldreich, P., \&
  Sridhar, S.\ 1995, \apj, 438, 763

\bibitem[Goodman \& Narayan(1989)]{goodman1989} Goodman, J., \&
  Narayan, R.\ 1989, \mnras, 238, 995

\bibitem[Greisen et al.(2009)]{greisen2009} Greisen, E.~W., Spekkens,
  K., \& van Moorsel, G.~A.\ 2009, \aj, 137, 4718

\bibitem[Gwinn et al.(1998)]{gwinn1998} Gwinn, C.~R., Britton, M.~C.,
  Reynolds, J.~E., Jauncey, D.~L., King, E.~A., McCulloch, P.~M.,
  Lovell, J.~E.~J., \& Preston, R.~A.\ 1998, \apj, 505, 928

\bibitem[Gwinn et al.(1991)]{gwinn1991} Gwinn, C.~R., Danen, R.~M.,
  Middleditch, J., Ozernoy, L.~M., \& Tran, T.~Kh.\ 1991, \apjl, 381,
  L43

\bibitem[Gwinn et al.(2014)]{gwinn2014} Gwinn, C.~R., Kovalev, Y.~Y.,
  Johnson, M.~D., \& Soglasnov, V.~A.\ 2014, \apjl{}L, in press,
  arXiv:1409.0530

\bibitem[H\"{o}gbom(1974)]{hogbom1974} H\"{o}gbom, J.~A.\ 1974, \aaps,
  15, 417

\bibitem[Jennison(1958)]{jennison1958} Jennison, R.~C.\ 1958, \mnras,
  118, 276

\bibitem[Johannsen \& Psaltis(2010a)]{johannsen2010a} Johannsen, T., \&
  Psaltis, D.\ 2010a, \apj, 716, 187

\bibitem[Johannsen \& Psaltis(2010b)]{johannsen2010b} Johannsen, T., \&
  Psaltis, D.\ 2010b, \apj, 718, 446

\bibitem[Joshi et al.(2009)]{joshi2009} Joshi, N., Zitnick, C.~L.,
  Szeliski, R., \& Kriegman, D.\ 2009, IEEE Conf.\ on Computer Vision
  and Pattern Recognition, 1550

\bibitem[Krishnan \& Fergus(2009)]{krishnan2009} Krishnan, D., \&
  Fergus, R.\ 2009, Neural Information Processing Systems, 1033

\bibitem[Lambert \& Rickett(1999)]{lambert1999} Lambert, H.~C., \&
  Rickett, B.~J.\ 1999, \apj, 517, 299

\bibitem[Levin et al.(2009)]{levin2009} Levin, A., Weiss, Y., Durand,
  F., \& Freeman, W.~T.\ 2009, IEEE Conf.\ on Computer Vision and
  Pattern Recognition, 1964

\bibitem[Levin et al.(2011)]{levin2011} Levin, A., Weiss, Y., Durand,
  F., \& Freeman, W.~T.\ 2011, IEEE Conf.\ on Computer Vision and
  Pattern Recognition, 2657

\bibitem[Lo et al.(1998)]{lo1998} Lo, K.~Y., Shen, Q.-Z., Zhao, J.-H.,
  \& Ho, P.~T.~P.\ 1998, \apjl, 508, L61

\bibitem[{\L}oza et al.(2009)]{loza2009} {\L}oza, A., Mihaylova, L.,
  Bull, D., \& Canagarajah, N.\ 2009, Machine Vision \& Applications,
  20, 71

\bibitem[Lu et al.(2011)]{lu2011} Lu, R.-S., Krichbaum, T.~P., Eckart,
  A., K\"{o}nig, S., Kunneriath, D., Witzel, G., Witzel, A., \&
  Zensus, J.~A.\ 2011, \aap, 525, A76

\bibitem[Lu et al.(2014)]{lu2014} Lu, R.-S., Broderick, A.~E., Baron,
  F., Monnier, J.~D., Fish, V.~L., Doeleman, S.~S., \& Pankratius,
  V.\ 2014, \apj, 788, 120

\bibitem[Mo\'{s}cibrodzka et al.(2009)]{moscibrodzka2009}
  Mo\'{s}cibrodzka, M., Gammie, C.~F., Dolence, J.~C., Shiokawa, H.,
  \& Leung, P.~K.\ 2009, \apj, 706, 497

\bibitem[Narayan \& Goodman(1989)]{narayan1989} Narayan, R., \&
  Goodman, J.\ 1989, \mnras, 238, 963

\bibitem[Noble et al.(2007)]{noble2007} Noble, S.~C., Leung, P.~K.,
  Gammie, C.~F., \& Book, L.~G.\ 2007, Classical and Quantum Gravity,
  24, 259

\bibitem[Reid et al.(2008)]{reid2008} Reid, M.~J., Broderick, A.~E.,
  Loeb, A., Honma, M., \& Brunthaler, A.\ 2008, \apj, 682, 1041

\bibitem[Rickett(1990)]{rickett1990} Rickett, B.~J.\ 1990, \araa, 28,
  561

\bibitem[Rogers et al.(1974)]{rogers1974} Rogers, A.~E.~E., et
  al.\ 1974, \apj, 193, 293

\bibitem[Russ(2011)]{russ2011} Russ, J.~C.\ 2011, The Image Processing
  Handbook (6th ed.; Boca Raton, FL; CRC Press)

\bibitem[Shen et al.(2005)]{shen2005} Shen, Z.-Q., Lo, K.~Y., Liang,
  M.-C., Ho, P.~T.~P., \& Zhao, J.-H.\ 2005, \nat, 438, 62

\bibitem[Spangler \& Gwinn(1990)]{spangler1990} Spangler, S.~R., \&
  Gwinn, C.~R.\ 1990, \apjl, 353, L29

\bibitem[Tatarskii(1971)]{tatarskii1971} Tatarskii, V.~I.\ 1971, The
  effects of the turbulent atmosphere on wave propagation

\bibitem[van Langevelde et al.(1992)]{vanlangevelde1992} van
  Langevelde, H.~J., Frail, D.~A., Cordes, J.~M., \& Diamond,
  P.~J.\ 1992, \apj, 396, 686

\bibitem[Wang et al.(2004)]{wang2004} Wang, Z., Bovik, A.~C., Sheikh,
  H.~R., \& Simoncelli, E.~P.\ 2004, IEEE Transactions on Image
  Processing, 13, 600

\bibitem[Wiener(1949)]{wiener1949} Wiener, N.\ 1949, The
  Extrapolation, Interpolation, and Smoothing of Stationary Time
  Series with Engineering Applications

\bibitem[Yan et al.(2014)]{yan2014} Yan, W., Vincent, F.~H.,
  Abramowicz, M.~A., Zdziarski, A.~A., \& Straub, O.\ 2014, \aap,
  submitted, arXiv:1406.0353

\bibitem[Zoran \& Weiss(2011)]{zoran2011} Zoran, D., \& Weiss,
  Y.\ 2011, IEEE International Conf.\ on Computer Vision, 479

\end{thebibliography}
\end{document}